# Single-Step Grayscale Lithography of Multi-Depth Mie Void Metasurfaces

*Oren Goldberg\*, Noa Mazurski and Uriel Levy*

Oren Goldberg[1], Noa Mazurski[1], and. Uriel Levy[1,2]
1 - Institute of Applied Physics, The Faculty of Science, The Center for Nanoscience and Nanotechnology, The Hebrew University of Jerusalem, Jerusalem 91904, Israel
2 - Singapore-HUJ Alliance for Research and Enterprise (SHARE), The Smart Grippers for Soft Robotics (SGSR) Programme, Campus for Research Excellence and Technological Enterprise (CREATE), Singapore 138602, Singapore
Corresponding author: Oren.goldberg@mail.huji.ac.il

**Abstract**
The height of dielectric metasurfaces is largely considered a constant in the fabrication process due to the top-down fabrication approach, resulting in a binary structure. Yet, for the recently introduced Mie voids metasurfaces, controlling the thickness of the voids locally is crucial for achieving significant spectral tuning. In this work we demonstrate Mie voids metasurfaces with local precise depth control using electron beam grayscale lithography. We underexpose PMMA with varying doses, which in turn translates to multiple depth levels in the developed resist. Transferring the pattern to a silicon substrate we generate Mie voids, trapping the light in the void which generates colors in reflection. By controlling the depth of the void at the nanoscale, we tune the resonance over the whole visible range and with high precision, resulting in a large gamut of colors, which is demonstrated with spectral measurements, images of uniform patterns and spatially varying patterns showcasing different geometrical designs and a detailed artistic image. The demonstrated approach can be used for the implementation of various types of dielectric metasurfaces, providing an additional important degree of freedom for their realization, with potential applications in structured light and structural colors, imaging, robotics, polarization control, sensing, virtual reality and more.

**Introduction**

Standard metasurface fabrication is inherently two-dimensional, as it typically relies on conventional top-down semiconductor nanofabrication processes[1–4]. In this approach, the out-of-plane (vertical) dimension is fixed at material deposition, while only the in-plane geometry is subsequently patterned. As a result, the structure is binary – at each lateral position the material is either present at its full height or absent. This constraint limits the range of metasurface designs that can be realized—particularly those relying on vertical degrees of freedom to control optical response. Metasurfaces can broadly be categorized as non-resonant or resonant structures [5–8]. Non-resonant devices rely on both lateral and vertical dimensions to accumulate phase[9–12]. Increasing height can reduce lateral footprint, thereby mitigating crosstalk between adjacent elements. Resonant metasurfaces, such as Mie-type structures, also benefit from vertical tunability: adjusting height modifies the scattering cross section without altering lateral spacing[13–15]. Finally, in quasi-bound-state-in-the-continuum (qBIC) metasurfaces, vertical symmetry breaking provides an alternative tuning knob to the more common in-plane perturbations—offering additional control over resonance properties[16–19]. While most of the resonant metasurfaces support resonances in high index materials, surrounded by a low index medium, the recently introduced Mie voids operated differently.

These Mie voids are a class of dielectric nano resonators in which light is trapped within an air-filled (or other low index media) cavity, surrounded by high-index material[20,21]. This confinement arises from strong internal reflection at the cavity walls, enabling strong local field enhancement. Such resonators have been demonstrated to support vivid structural colors and high refractive-index sensitivity[20,21].

Several methods can be used to remove the dimensionality limitation along the vertical dimension. One notable example is that of a multi-layer device[22–25]. Another possible way is to use grayscale lithography. Several grayscale lithography approaches have been explored to enable height control in micro- and nanophotonic devices[26–30].
These include grayscale photolithography using attenuated masks, nanoimprint lithography with relief masters, focused ion beam (FIB) milling[20,31], and dose-modulated electron-beam lithography[32,33]. While each technique provides vertical structuring to some extent, limitations such as diffraction-limited resolution, low throughput, or restricted scalability have hindered their adoption for nanoscale metasurface platforms[34–36].

Here, we introduce an electron-beam grayscale lithography (eBGSL) method for achieving vertical tunability in dielectric metasurfaces. By locally varying the electron dose in a positive resist (PMMA), we induce partial chain scission, leaving a residue that modulates etch resistance. This enables precise depth control at the nanoscale across the metasurface in a single fabrication step, with local etch rate modulation determined by the spatial dose profile. The resonant wavelength of the Mie void is highly sensitive to the depth of the void—making it ideally suited to grayscale-based fabrication. In their original demonstration[20], both the lateral and vertical dimensions of Mie voids were defined using focused ion beam milling—a slow, serial process that is unsuitable for large-area fabrication. In contrast, the eBGSL method introduced here enables scalable, parallel definition of depth-controlled voids with subwavelength resolution and full design flexibility.

**Results and Discussion**
We first study numerically the reflection spectra of arrays of holes in silicon with varying radii and depths using a commercially available RCWA software (Ansys inc, see the methods section for further details on the simulation parameters). The reflection spectra are converted to values on the CIE 1937 chromaticity diagram using color matching function[37]. Fig 1.a,b represents the same data from the simulations but colored differently. In Fig. 1a, radius is fixed while depth varies. In Fig. 1b, depth is fixed while radius varies. When the radius is held constant and the void depth is varied, the resulting reflection colors span a wide, continuous region of the CIE color space (see trajectory for a given color of the dots in the image). In contrast, varying the radius at fixed depths yields a more localized and limited color range. This is also illustrated in fig S1. It can be clearly seen that depth control is the critical factor for tunability in Mie void metasurfaces. As depth control is not compatible with standard top-down procedures, there is a need for a process which has control over the depth in addition to in plane dimensions. Electron beam gray scale lithography (eBGSL) is a process which addresses these needs, as the depth is controlled by the electron beam dosage and the mask transfer to the substrate is done in parallel in the etching stage, as shown schematically in fig 1.c. As such, we explore eBGSL using a positive eBeam resist (PMMA) which has been shown to have a gradual decreasing profile[38,39] allowing for a broad range of resist thicknesses across varying doses.
We validated this concept by spin coating a 650nm thick layer of PMMA on a silicon substrate and exposed arrays made of holes with a radius of 340nm and a pitch of 1$\mu$m. A wide range of doses was used to cover the whole range from complete underexposure to overexposure. After development, the sample was measured using atomic force microscopy to determine the remaining PMMA thickness at the exposed locations (see fig S2 for selected measurements). The results, shown in Fig. 1d, reveal a monotonic increase in depth with increasing doses. A light-orange trend line qualitatively highlights four distinct regimes in the dose–depth response: a gradual onset, a linear region, a sharp transition, and saturation

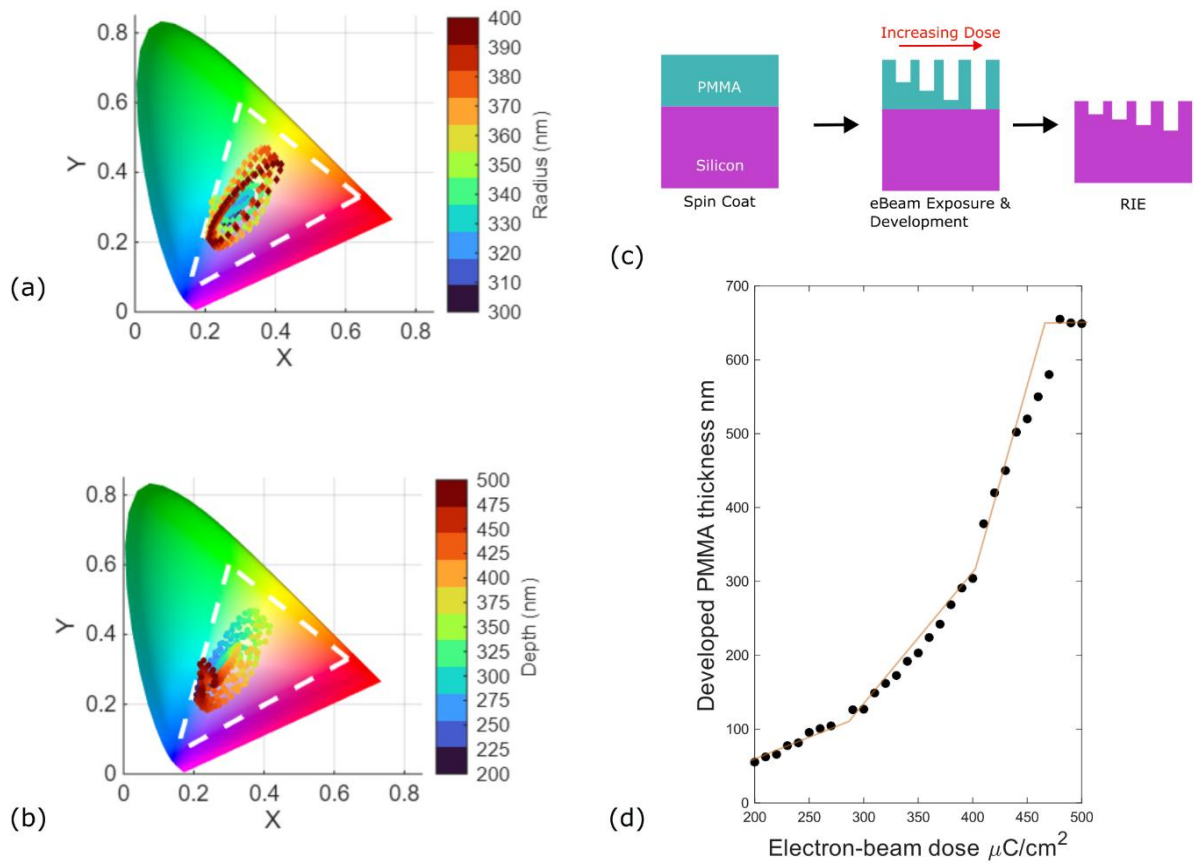

*Figure 1: a,b) Simulated reflection spectra of Mie-void metasurfaces with varying radii and depths, plotted on the CIE 1931 chromaticity diagram. The data shown in both panels are identical; in (a), points corresponding to the same radius are assigned to the same color, while in (b), points corresponding to the same depth are assigned the same color. It can be seen that for dots with the same color, the range of color tuning in a is significantly larger than the range in b, indicating that depth variation can provide a significantly larger color tuning. c) Fabrication schematic illustrating the single-step grayscale lithography process enabling multi-depth etching. d) Post-exposure developed PMMA thickness as a function of electron-beam dose. Qualitive trend of the graph is indicated by light orange line.*

It can be clearly seen that a large depth profile is achieved over a controlled dose region. Following these results, we transferred the patterned resist to the substrate using a reactive ion etching (RIE). Under the etch conditions (detailed in the methods section), the etch rates of PMMA and silicon were measured to be 1.15nm/s and 0.7nm/s, respectively, yielding a selectivity of approximately 1.64:1 (PMMA:silicon). It is important to note that the final depth is highly dependent on the specific conditions of the RIE process. As such, the etching process needs to be calibrated carefully to achieve the desired depth profile, for further details see methods section.

Based on simulations, we found that spanning the entire visible spectrum requires a PMMA layer of 800nm thickness. As such, another sample was prepared and etched as described before. Optical microscope images of the Mie void metasurface in silicon are shown in fig. S3 spanning a wide color gamut. The voids are formed uniformly as seen in fig S4 and a depth difference between the arrays is demonstrated in fig. S5.

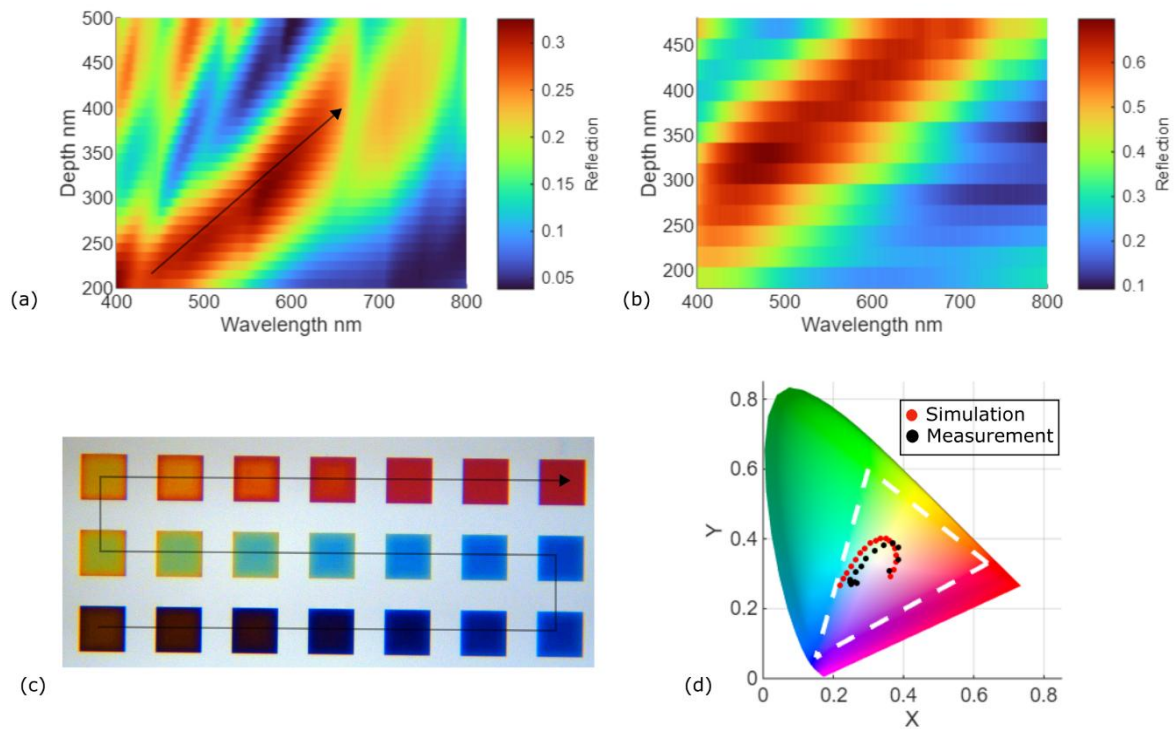

*Figure 2: Comparison between simulated and measured Mie void metasurfaces. a,b) Simulated (left) and measured (right) reflection spectra of depth varying Mie void metasurfaces. Measured spectra are referenced to silicon substrate. c) Optical microscope image of the sample from (b). Black line indicates the direction of increasing depth. d) Comparison of simulated and measured spectra on the CIE 1931 chromaticity diagram. Dashed white line represents the sRGB.*

Reflection measurements were performed and compared to simulations, as seen in fig. 2a, b. We see another confirmation to the observation that the Mie void resonance is dependent on the depth of the void and is tunable over the whole visible spectrum. Measured and simulated reflection spectra are in good agreement, although the measured spectra exhibit higher reflectivity, due to referencing to un-patterned silicon substrate as appose to air. Fig 2.c is an optical microscope image of the measured arrays, faded black line represents the increasing depth trend giving us visual confirmation that a multi-depth etch was achieved. We compare the simulated and measured spectra on the chromaticity diagram and see a very good correlation between the two. We attribute the discrepancies to the deviation of the hole profile from a perfect cylinder at the rim, as seen in fig S4.

So far, we've demonstrated that we can perform a multi-depth etch in a single cycle, reducing the time consumption in the process, but furthermore this process does not need an alignment step which naturally introduces difficulties into the fabrication process. To further demonstrate the synergetic relationship the eBGSL has with the Mie void platform we designed several intricate patterns with varied spatial features demonstrating that the lithography process isn't limited to uniform patterns.

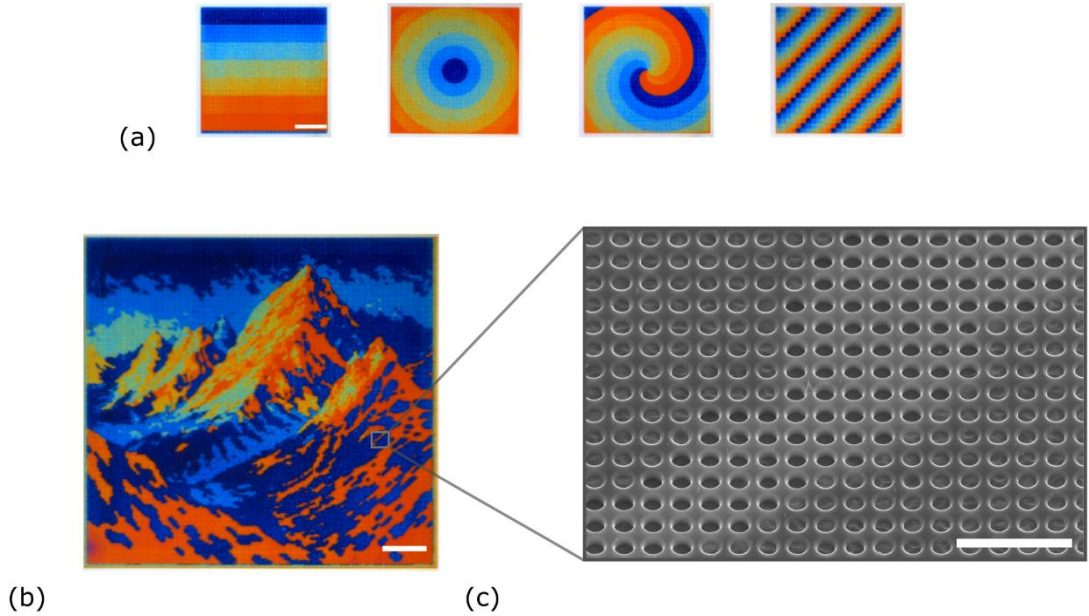

*Figure 3: a) Optical microscope images of metasurfaces patterned with spatially varying, depth-controlled profile; scale bar is 50μm. b) Large-area optical microscope image of a fabricated metasurface with spatially varying depth profiles forming an artistic pattern; scale bar is 50μm. c) SEM image of a magnified region from (b), confirming uniform periodic arrays of air voids with distinct depth variations; Scale bar is 4μm.*

We repeated the process as before and coated a silicon substrate with 800nm of PMMA. We then designed several patterns with spatially varying features such as bullseye, gradient spiral and mosaic where seven grayscale representative levels were chosen (see fig S6.a) to showcase that this platform is not limited to only wide area periodic designs. The optical images of fabricated samples can be seen in fig 3.a. We further demonstrate spatially resolved coloration by fabricating a complex artistic image—a sunset-lit mountain range—encoded through locally varying void depths (target images can be seen in fig S6.b,c). The resulting image can be seen in fig 3.b, showcasing detailed intricate coloring, with close agreement with the target image. Fig 3.c depicts a close-up SEM of the fig 3.b, where the color difference correlates to depth differences in the Mie voids.

In summary, we showcase a platform which enables a single-step grayscale lithography for a multi depth metasurface. We first demonstrate that the eBGSL can be achieved by underexposing PMMA getting a wide-ranging depth values for a large dose range enabling accurate depth control. We show that specifically Mie void resonators benefit tremendously from this platform as the resonance wavelength is mostly dependent on the depth of the void rather than the diameter (which is not the case for many resonative dielectric metasurfaces). Subsequently, samples were prepared with varying depths. Spectral measurements were in good agreement with simulations and optical images verified by the appearance of structural colors. We further demonstrated that this platform is not limited to uniform patterns and fabricated spatially varying samples showcasing different geometrical designs and a detailed artistic image. This platform opens new avenues for depth dependent resonative and non-resonative dielectric metasurfaces, with myriad potential applications such as structured light and structural colors, imaging, robotics, polarization control, sensing, virtual reality and more.

**Methods**
Sample preparation:
A bare silicon substrate 20X20mm is prebaked on a hot plate at 180c for 5 min. The silicon is then spin coated with PMMA 495A for 50sec at 2300 rpm and baked for 2 min on a hot plate.

This is then repeated, resulting in a ~800nm thick PMMA thin film (400nm for each coating). We then write on the sample using E-ebeam lithography (Elionix ELS-G100) at a current of 1nA. The dose test for the arrays was done using an internal function of the eBeam software. For spatially varying samples we used BEAMER software (GenISys) to distribute the dose to the different spatial positions. We used proximity effect correction to compensate for spatially varying doses. The sample was developed in a MIBK:IPA (1:3) solution for 1 min, and then subsequently rinsed in IPA for 10 sec which was then dried using N2. We used reactive ion etching (Corial 2001) to etch the sample with an etching recipe consisting of SF6 and CH4. Before etching, we conditioned the chamber first running it on empty and then again with an un-patterned sample (PMMA on silicon) for 200 sec. The etched un-patterned sample is then used to measure the etching rate of the PMMA. The PMMA thickness is measured with ellipsometry (J.A. Woollam). The etching time is calculated according to the etch rate and thickness of the PMMA.

Optical measurement:
Spectral measurements were performed using an inverted microscope Nikon Eclipse TE300, the sample is illuminated with a tungsten-halogen 100W lamp. The light is focused on to the sample via a 10X objective. The same objective collects the reflected light which is then coupled to a spectrometer via a 50mm doublet lens (ThorLabs AC254-050-AB) and multimode optical fiber (Ocean Insight QP600-2-VIS-NIR).
The optical images were captured using an optical microscope Vickers Instruments Compound Binocular Microscope, the sample is illuminated with a halogen lamp. The light is focused on the sample via objective (OLYMPUS SLMPLN 20x) and collected via the same objective. The reflected light is imaged on a CMOS camera (ImagingSource DFK 33UX183).

Simulations:
Simulations were performed using the commercially available RCWA software (Ansys Lumerical). All simulations consist of a silicon substrate (optical constants taken from in software library), where the void was defined as an etch region with ranging radius R=300-400nm, and depth of 200-500nm. The period was kept constant throughout all simulations at 1 $\mu$m. We use the "grating power" result and extract only the zero order reflection.

Research funding: This research is partially supported by the Israeli innovation authority within the framework of the Israeli meta-materials and Meta-surfaces consortium. This research is partially supported by a grant from the National Research Foundation, Prime Minister's Office, Singapore under its Campus of Research Excellence and Technological Enterprise (CREATE) program.